\documentclass{qjmam}

\startpage{1}
\yr{2016}
\vol{0}
\issue{0}

\usepackage{amsmath,amsthm,amssymb}
\usepackage{comment}
\usepackage[a4paper]{geometry}
\usepackage{appendix}

\DeclareMathOperator{\Div}{div}

\newcommand{\R}{\mathbb{R}}
\newcommand{\pd}[2]{\frac{\partial #1}{\partial #2}}

\newtheorem{theorem}{Theorem}

\everymath{\displaystyle}

\DeclareMathAlphabet\mathbit
    \encodingdefault\rmdefault\bfdefault\itdefault
\DeclareOldFontCommand{\bi}{\normalfont\bfseries\itshape}{\mathbit}

\def\fakebold#1{\relax\ifvmode\leavevmode\fi%
\ifmmode%
\setbox0=\hbox{$#1$}%
\else%
\setbox0=\hbox{#1}%
\fi%
\kern-.02em\copy0 \kern-\wd0%
\kern .04em\copy0 \kern-\wd0%
\kern-.0125em\raise.02em\box0%
}%

\begin{document}

\title[existence~of~self-similar~converging~shocks~for~arbitrary~EOS]{ON THE EXISTENCE OF SELF-SIMILAR CONVERGING SHOCKS FOR ARBITRARY EQUATION OF STATE}
\author[boyd,~ramsey,~and~baty]{ZACHARY M. BOYD,\footnote{zach.boyd@math.ucla.edu}}
\def\lanl{Los Alamos National Laboratory,
New Mexico, United States}
\def\ucla{Mathematics Department, UCLA, Los Angeles, California, United States}
\address{\lanl \\
\and \ucla}
\extraauthor{SCOTT D. RAMSEY,\footnote{ramsey@lanl.gov} \and ROY S. BATY\footnote{rbaty@lanl.gov}}
\extraaddress{\lanl}
\received{\recd 6 October 2016.}

\maketitle

\eqnobysec

\begin{abstract}
We extend Guderley's problem of finding a self-similar scaling solution for a converging cylindrical or spherical shock wave from the ideal gas case to the case of flows with an arbitrary equation of state closure model, giving necessary conditions for the existence of a solution. The necessary condition is a thermodynamic one, namely that the adiabatic bulk modulus, $K_S$, of the fluid be of the form $pf(\rho)$ where $p$ is pressure, $\rho$ is mass density, and $f$ is an arbitrary function. Although this condition has appeared in the literature before, here we give a more rigorous and extensive treatment. Of particular interest is our novel analysis of the governing ordinary differential equations (ODEs), which shows that, in general, the Guderley problem is always an eigenvalue problem. The need for an eigenvalue arises from basic shock stability principles -- an interesting connection to the existing literature on the relationship between self-similarity of the second kind and stability. We also investigate a special case, usually neglected by previous authors, where assuming constant shock velocity yields a reduction to ODEs for every material, but those ODEs never have a bounded, differentiable solution. This theoretical work is motivated by the need for more realistic test problems in the verification of inviscid compressible flow codes that simulate flows in a variety of non-ideal gas materials.
\end{abstract}

\section{Introduction}

While much is understood concerning inviscid compressible flows in ideal gases, less is known about counterpart flows in non-ideal materials. For instance, in explosion and high-speed impact phenomena, metals, plastics, and other non-gaseous materials experience pressures far exceeding the limits of stress-strain models and are instead well-modeled by inviscid compressible flow (Euler) equations. However, because of the structure of these materials, constituitive relations other than the ideal gas law are needed to correctly model the thermodynamics involved. Some equations of state (EOS) that find use in these situations are very dissimilar to that of an ideal gas and may only be valid in a limited range of pressures, energies, and densities. These EOS models can arise from theoretical considerations, empirical measurements, or a combination of both. Understanding these types of flows on a theoretical and computational level leads to improvements in safety and reliability in explosives handling, aids in the design of blast-resistant materials, and enables enhanced experimental design.

Inviscid compressible flow codes (or ``hydrocodes'') are widely used to simulate explosive or high-speed impact phenomena, and must be subject to rigorous programs of verification and validation (as reviewed among many others by Oberkampf et al.~\cite{oberkampf2004verification,oberkampf2010verification}, Roy~\cite{roy2005review}, and Kamm~\cite{oldies_but_goodies}) while being so employed. One important code verification tool is comparison to exact solutions of the underlying equations otherwise solved approximately by a hydrocode. In the ideal gas case the Noh~\cite{noh,rider,gehmeyr,our_shock_waves_paper}, Sedov \cite{sedov_book}, Guderley~\cite{guderley,lazarus,guderley_revisited}, Kidder~\cite{kidder1974theory}, and Coggeshall~\cite{coggeshall1986lie,coggeshall1991analytic,coggeshall1992group} solutions are example ``test problems'' that may be used to reveal strengths and weaknesses in a hydrocode (e.g., wall-heating errors, symmetry breaking). Unfortunately, in the case of non-ideal gas flows, exact solutions are at best less well-understood, if not altogether unavailable.

There has been quite a bit of effort to export some of the aforementioned ideal gas solutions to the inviscid compressible flow of non-ideal materials, mostly by identifying EOS models with special properties and mimicking the derivations from the corresponding ideal gas problems~\cite{wu_roberts,jena_sharma,tait,abel_noble,noh_matls}. There have also been some attempts at treating these issues in generality~\cite{holm1976symmetry,axford_holm_1978,hutchens,our_shock_waves_paper}, some of which have been successful. On the other hand, it has become clear that there is something special about the ideal gas constitutive law that allows for the existence of more exact solutions than other, more realistic EOS models: namely, a lack of inherent dimensional scales in both the Euler equations and EOS, and a corresponding abundance of symmetries. Thus, for each problem, it seems that the best we can do is find all EOS models that possess the necessary symmetries for the Euler equations to have exact solutions with desired properties. This method essentially amounts to finding forms of an EOS that lead to a reduction of the Euler equations from partial differential equations (PDEs) to ordinary differential equations (ODEs), and then testing whether the ODEs have a solution with desirable properties. While the reduction to ODEs has been well-studied, the solvability of those ODEs has received less attention (with the notable exceptions of Guderley~\cite{guderley}, Sedov~\cite{sedov_book}, and Lazarus~\cite{lazarus}). One of the contributions of this work is to initiate that study in an important special case. Existence in these problems is not trivial and often reduces to a nonlinear eigenvalue problem, whose solution properties appear to not be rigorously understood.

In addition to the need for useful code verification problems, there are two other motivations for finding analytical solutions for non-ideal, inviscid compressible flows. The first -- discussed principally by Barenblatt~\cite{barenblatt} -- is that such solutions frequently express the intermediate asymptotic behavior of physical systems. Additional effort has gone into the closely related problem of showing that different kinds of exact solutions are stable to perturbation and in some sense will attract nearby solutions to them~\cite{ponchaut_2005,ponchaut_2006,hornung2008question,morawetz,hafele,brushlinskii,clarisse}.

Our final motivation is the fact that in the context of inviscid compressible flow, many exact solutions belong to the sub-classes of self-similar or scale-invariant solutions. Understanding the spatial, temporal, or other scaling behavior of fluid flow scenarios allows experimentalists to choose -- ostensibly without penalty -- the dimensional scale on which to perform work, and then extrapolate according to simple scaling laws. This is the main idea behind ``scale modeling'' in engineering, and it often helps achieve significant cost savings, increases the amount of data that can be obtained, and in some cases even makes the difference between an experiment being feasible or not.

\subsection{Contributions of this work}

In this work, we focus on the Guderley idealized implosion problem~\cite{guderley,lazarus,meyer_ter_vehn_schalk,chisnell,guderley_revisited}, which considers an infinitely strong cylindrically or spherically symmetric shock wave moving toward the one-dimensional curvilinear origin, focusing at the origin, and then reflecting back into the surrounding once-shocked medium. In addition to being solved by the aforementioned authors in the ideal gas case, the Guderley problem has also been solved in some other scenarios~\cite{wu_roberts,jena_sharma,tait}, and Holm~\cite{holm1976symmetry}, Axford and Holm~\cite{axford_holm_1978}, and Hutchens~\cite{hutchens} even provide a general class of EOS models for which there exists a reduction from the PDEs of inviscid compressible flow to ODEs. For the purposes of code verification, a reduction to ODEs is often considered an ``exact'' solution because numerical ODE solvers are usually considered to be more robust than the hydrocodes (in general, numerical PDE solvers) they are often used to verify.

Moreover, Ramsey et al.~\cite{our_shock_waves_paper} have shown that certain classical ideal gas test problems can be solved for essentially arbitrary EOS closures, using a choice of similarity variables and associated reduction-to-ODEs that applies equally well regardless of the material in which the compressible flow occurs. On the other hand, there is an obstacle in the case of curvilinear flows that prevents the existence of counterpart universal solutions. For example, Ramsey et al.~\cite{our_shock_waves_paper} find that the one-dimensional cylindrical or spherical Noh problems only have solutions for certain choices of EOS. Such issues with curvilinear symmetry do not arise because of incompatible boundary conditions, but rather from the fact that the reduced ODEs do not have a non-trivial solution with the properties that define a generalized Noh flow.

Indeed, while the presence of enough symmetries to reduce the governing Euler PDEs to ODEs has been extensively studied (see, for example, Ovsiannikov~\cite{ovsiannikov_book}, Andreev et al.~\cite{andreev1998applications}, Cantwell~\cite{cantwell}, Holm~\cite{holm1976symmetry}, Axford and Holm~\cite{axford_holm_1978}, and Hutchens~\cite{hutchens}), less is known about when the resulting ODEs actually have a solution that satisfies conditions specific to a given class of problems. In this work, we obtain some first results concerning the aforementioned ODEs, which arise from considering the Guderley problem. We also hope that our analysis will provide convincing evidence that the existence of suitable solutions is a non-trivial consideration, which needs to be explicitly addressed in studies involving similarity solutions, especially when working with problems exhibiting Zel'dovich and Raizer's and Barenblatt's ``self-similarity of the second kind''~\cite{zeldovich_and_raizer,barenblatt}.

This paper thus contributes to the literature in at least the following ways:
\begin{itemize}
\item Holm~\cite{holm1976symmetry}, Axford and Holm~\cite{axford_holm_1978}, and Hutchens~\cite{hutchens} derivation of a class of EOS closure models in which the Guderley problem has an ODE solution is made more rigorous and complete.
\item The unshocked conditions that permit the existence of a Guderley solution are treated in more detail than in previous works, as far as we are aware. This treatment highlights some counterintuitive facts about the existence of ``strong shocks'' and self-similar scaling in the case where there exist characteristic unshocked density, pressure, and specific internal energy scales. These results seem to contradict, or at least require a more subtle use of, the rule of thumb commonly employed to determine the number of self-similar scalings admitted by a problem formulation.
\item The ODEs that result from the introduction of self-similar scaling variables are treated in full generality.
\item The important special case of a universal reduction to ODEs (i.e., the universal symmetry as noted by Ovsiannikov~\cite{ovsiannikov_book}, Holm~\cite{holm1976symmetry}, and Boyd et al.~\cite{general_euler_symmetries}) is treated. We show that in this case, the associated ODEs never have a bounded Guderley solution. This result motivates further study of the ODEs resulting from symmetry or scaling reduction, rather than concentrating solely on the symmetries.
\item We show that with the standard choice of scale-invariant similarity variables, in a Guderley scenario one must choose between an eigenvalue problem and unbounded solutions.
\item We provide a novel intuitive explanation for the blowup in the ODE solve that occurs for almost all choices of similarity variables: there is a competition between boundedness and shock stability which underlies the blowup.
\end{itemize}

In support of these objectives, Sec.~\ref{sec:math}, includes an overview of the inviscid Euler equations, a rigorous definition of the Guderley problem, and a discussion of the permissible unshocked fluid states for which this problem may be defined. In Sec.~\ref{sec:main}, we include a formalized proof surrounding the existence of bounded Guderley solutions for a general EOS closure model. We conclude in Sec.~\ref{sec:conclusion}.

\section{Mathematical model}
\label{sec:math}

The equations of adiabatic motion for an inviscid compressible fluid are given by (see, e.g., Landau and Lifschitz~\cite{landau_and_lifschitz}, Courant and Friedrichs~\cite{courant_and_friedrichs}, Ovsiannikov~\cite{ovsiannikov_book}, Axford~\cite{axford}, and Harlow and Amsden~\cite{harlow1971fluid})
\begin{eqnarray}
	d_t\rho + \rho \Div \bf{u} = 0, \label{eqn:euler_3d_begin}\\
	\rho d_t \bf{u} + \nabla p = 0, \\
	d_t p + K_S \Div \bf{u} = 0,
  \label{eqn:euler_3d_end}
\end{eqnarray}
where 
\begin{equation}
	d_t = \partial_t + \bf{u} \cdot \nabla, 
\end{equation}
is the material derivative, $u$ is the velocity field, $p$ is the (scalar) pressure, $\rho$ is the mass density, and $K_S$ is the \emph{adiabatic bulk modulus}, defined as
\begin{equation}
  K_S \equiv \rho\left( \pd{p}{\rho} \right)_S,
  \label{eqn:abm_def}
\end{equation}
where $S$ is the fluid entropy. The adiabatic bulk modulus is also related to the local speed of sound $c$ by 
\begin{equation}
  K_S = \rho c^2,
  \label{eqn:sound_speed_def}
\end{equation}
and is in general a material-dependent function of $p$ and $\rho$ -- indeed, the introduction of the adiabatic bulk modulus is the only source of information about the thermodynamic properties of the specific material under consideration. Moreover, given an EOS of the form $p = p(e,\rho)$ (where $e$ is the energy per unit mass or specific internal energy; SIE), $K_S$ is obtained from the relation 
\begin{equation}
  K_S=\rho \left.\pd{p}{\rho}\right|_e+\left.\frac{p}{\rho}\pd{p}{e}\right|_\rho.
  \label{eqn:bmdef}
\end{equation}
For example, for the ideal gas EOS,
\begin{equation}
  p=(\gamma-1) \rho e,
  \label{eqn:ideal_gas}
\end{equation}
the corresponding adiabatic bulk modulus is determined from Eq.~(\ref{eqn:bmdef}) to be
\begin{equation}
  K_S=\gamma p, 
  \label{eqn:bmideal}
\end{equation}
where $\gamma > 1$ is the polytropic index. In general, the reverse relation
\begin{equation}
  \left.K_S\pd{e}{p}\right|_\rho+\left.\rho\pd{e}{\rho}\right|_p = \frac{p}{\rho},
  \label{eqn:Ks_invert}
\end{equation}
shows that an inverted EOS of the form $e = e(p,\rho)$ can also be recovered from $K_S$ (as shown in detail by Axford~\cite{axford}). 

In any event, Eqs.~(\ref{eqn:euler_3d_begin})-(\ref{eqn:euler_3d_end}) express the conservation of mass, momentum, and energy of the fluid, neglecting heat conduction, body forces, material strength, anisotropy of material structure, and viscosity. These assumptions are appropriate in many contexts where shock waves form, including supersonic flows~\cite{courant_and_friedrichs}, explosions~\cite{zeldovich_and_raizer}, shock tube experiments~\cite{zeldovich_and_raizer}, and space reentry~\cite{nasa_tables}.

Restricting to one-dimensional (1D) symmetry (which will be employed throughout the remainder of this work), Eqs.~(\ref{eqn:euler_3d_begin})-(\ref{eqn:euler_3d_end}) become
\begin{eqnarray}
  \pd{\rho}{t} + u \pd{\rho}{r} + \rho\left( \pd{u}{r} + \frac{k u}{r} \right) = 0, \label{eqn:euler_spherical_begin}\\
  \pd{u}{t}    + u \pd{u}{r}    + \frac{1}{\rho}\pd{p}{r} = 0, \\
  \pd{p}{t}    + u \pd{p}{r}    + K_S\left( \pd{u}{r} + \frac{k u}{r} \right) = 0,
  \label{eqn:euler_spherical_end}
\end{eqnarray}
where $k=1$ or $2$ for cylindrical or spherical symmetry, respectively. The planar symmetry case ($k = 0$) will not be considered in this work, as its properties are very different, and a wide variety of solutions can be shown to exist.

\subsection{Guderley's problem}
\label{sec:gud_prob}

We seek solutions to a natural extension of Guderley's idealized implosion problem~\cite{guderley}, which is a cylindrically or spherically symmetric converging-reflecting shock solution of Equations~(\ref{eqn:euler_spherical_begin})-(\ref{eqn:euler_spherical_end}). For the case of an ideal gas [i.e., $K_S$ given by Equation~(\ref{eqn:bmideal})], this problem has also been solved by Stanyukovich~\cite{stanyukovich}, Butler~\cite{butler}, Lazarus~\cite{lazarus}, and many others. Some variations on this problem have also been considered and solved~\cite{wu_roberts,jena_sharma,axford_holm_1978,tait}. In this work, we develop the theory of existence of Guderley solutions in materials with arbitrary $K_S$. 

The objective of the Guderley problem is to determine the flow in a scenario where a strong cylindrically or spherically symmetric shock generated far from the origin converges at the origin and is reflected back. Following authors such as Stanyukovich~\cite{stanyukovich}, Zel'dovich and Raizer~\cite{zeldovich_and_raizer}, and Chisnell~\cite{chisnell}, in this work we restrict to the converging regime. As shown in detail by Lazarus~\cite{lazarus} and Ramsey et al.~\cite{guderley_revisited}, analysis of the reflected regime follows naturally from that of the converging regime, and will be deferred to a future study.

For the sake of mathematical precision, we define the Guderley problem as the problem of finding functions $\rho,u,p$, and $r_s$ (the shock wave location as a function of $t$) satisfying the following requirements:
\begin{itemize}
\item $r_s:(-\infty,0]\to [0,\infty)$ and $\lim_{t\to -\infty} r_s(t)=\infty,$ $r_s(0)=0$. $r_s$ is monotonic. The time $t=0$ corresponds to the moment of focusing, so $t<0$ throughout this work.
\item $\rho,u,p:(-\infty,0)\times (0,\infty)\to \R$ are solutions to Eqs.~(\ref{eqn:euler_spherical_begin})-(\ref{eqn:euler_spherical_end}), and are differentiable except at $r_s(t),$ where a shock occurs.
\item $u$ and $p$ are negligibly small for $r<r_s$. We set them formally to zero. The unshocked SIE takes whatever value is determined by our choice of $\rho$ and $p$, and the EOS closure model. (See Sec.~\ref{sec:unshocked} for the case where $p=0$.) Density $\rho$ is a prescribed constant $\rho_0>0$ for $r<r_s(t)$.
\item $\rho$, $u$ and $p$ are bounded at each fixed $t<0$.
\item $K_S:[0,\infty)\times [0,\infty)\to [0,\infty)$ is bounded and continuous, and $K_S>0$ whenever $\rho>0$ and $p>0$.
\item $\rho>0$, $p>0$, and $u<0$ for $r>r_s(t)$.
\item The shock travels subsonically relative to a particle immediately behind it. This means that $(u_s-u_p)^2 < c^2$ at the coordinate $r = r_s + \epsilon$, where $\epsilon$ is small; here, $u_s$ and $u_p$ are the shock and aforementioned particle velocities, respectively. This is a stability condition and is implied by either of the following two common assumptions:
\subitem \emph{Thermodynamic stability}: $\left.\frac{\partial^2 p}{\partial\rho^2}\right\vert_S>0$, which is generally true of real physical systems away from phase transitions and assuming a compression rather than rarefaction shock. It arises from the more general condition that entropy should increase across the shock front, as discussed by Courant and Friedrichs~\cite{courant_and_friedrichs}.
\subitem \emph{Perturbative stability}, also known as evolutionary stability: if the unshocked and shocked regions are perturbed by the addition of small acoustic waves, the shock will remain a shock over time and only adjust in speed to accommodate the perturbations. This prevents the shock from devolving into a rarefaction, as discussed by Jeffrey~\cite{jeffrey} and Burgess~\cite{burgess}.
\end{itemize}

The Guderley solution is an example of a \emph{self-similar scaling solution} of Eqs.~(\ref{eqn:euler_spherical_begin})-(\ref{eqn:euler_spherical_end}), of the form
  \begin{eqnarray}
    \rho(r,t) &=& |t|^{\beta_{\rho}} D(\xi), \label{eqn:sim_ansatz_begin} \\
    u(r,t) &=& |t|^{\beta_u} V(\xi), \\
    p(r,t) &=& |t|^{\beta_p} \Pi(\xi), \label{eqn:sim_ansatz_end}
  \end{eqnarray}
  where 
  \begin{equation}
    \xi = r |t|^{-(1-\lambda)},
    \label{eqn:sim_ind} 
  \end{equation}
$D,V,$ and $\Pi$ are functions solely of the indicated argument, and $\beta_{\rho},$ $\beta_{u}$, $\beta_p$, and $\lambda$ are constants.
The constant $\lambda$ is referred to as the \emph{similarity exponent}, and it must satisfy $\lambda\in (0,1]$ as will be shown in Section~\ref{sec:shock_trajectory}. The constants appearing in Eqs.~(\ref{eqn:sim_ansatz_begin})-(\ref{eqn:sim_ansatz_end}) must be chosen so that Eqs.~(\ref{eqn:euler_spherical_begin})-(\ref{eqn:euler_spherical_end}) reduce to ODEs in the independent variable $\xi$ when this ansatz is assumed; whether or not this is possible depends on the symmetries present in the coupled PDE-EOS system.

There are other, equivalent similarity variable constructions; for example,
  \begin{align}
    \rho &= \rho_0 D(\xi,) \nonumber \\
    u &= \frac{r}{t}V(\xi), \nonumber \\
    p &= \rho_0 \frac{r^2}{t^2} \Pi(\xi), \nonumber \\
    \xi &= r (-t)^{\alpha}. \nonumber 
  \end{align}
One can also replace the pressure or adiabatic bulk modulus with the local sound speed, etc. -- indeed, many equivalent forms of the Euler equations and their similarity variables exist. The form of the Euler equations used in this work is identical to that used by, for example, Axford~\cite{axford} and Ovsiannkov~\cite{ovsiannikov_book}; the associated choice of similarity variables is that which arises most naturally from that form of the equations. Our form also appears naturally when one uses the symmetry analysis method~\cite{ovsiannikov_book,bluman_book,cantwell,olver}, which is a generalization of the approach used here.

From the standpoint of dimensional analysis~\cite{zeldovich_and_raizer}, each set of constitutive units present in the problem (e.g., meters, seconds, kilograms) adds one degree of scaling symmetry, and each inherent scale present in the problem (e.g., dimensional constants such as the speed of light) removes one degree of scaling symmetry.\footnote{This is not rigorously true, however. For example, Eqs.~(\ref{eqn:euler_spherical_begin})-(\ref{eqn:euler_spherical_end}) also admit symmetries that need not arise from dimensional considerations alone. One such symmetry is discussed in more detail in Sec.~\ref{sec:lambda_zero}.} The reasons for restricting to scaling solutions -- a very narrow class of symmetries -- are twofold: first, the ideal gas and other solutions are of this form, Second, such solutions can be understood by taking a single ``snapshot'' in time -- the solution at earlier or later times is just a scaled version of this snapshot.

Thus, for code verification and experimental design purposes, self-similar scaling solutions are particularly useful, being, in a sense, of reduced dimensionality compared to that of the governing equations under consideration. As discussed by Barenblatt~\cite{barenblatt}, they also tend to correspond to ``intermediate asymptotic'' solutions that physical flows approach when they are no longer strongly influenced by their initial conditions, thus yielding intuition about real physical flows -- although a separate stability analysis would be necessary to confirm that this connection is valid.

\subsection{The unshocked state}
\label{sec:unshocked}

The definition of the Guderley problem provided in Section~\ref{sec:gud_prob} includes the typical choice of unshocked state, in particular featuring $\rho=\rho_0$ and $u=0$. Since the problem definition also includes the presence of a shock wave (interpreted as a mathematical discontinuity in the context of the inviscid Euler equations), the Rankine-Hugoniot shock jump conditions must be employed at $r=r_s$, in lieu of Eqs.~(\ref{eqn:euler_spherical_begin})-(\ref{eqn:euler_spherical_end}) (from which they may be formally derived). These conditions express conservation of mass, momentum, and energy across the shock wave, and the form corresponding to the Guderley problem definition is given by
\begin{align}
  \rho_1(u_s-u_1) &= \rho_0 u_s, \label{eqn:jump_begin} \\
  \rho_1 u_s u_1 &= p_1-p_0, \\
  \rho_0 u_s \left(e_1-e_0 + \frac{1}{2}u_{1}^2\right) &= p_1 u_1, \label{eqn:jump_end}
\end{align}
where the subscripts $0$ and $1$ indicate unshocked and shocked quantities, and $u_s$ is again the shock velocity. 

The ideal gas Guderley problem includes $p_0 = e_0 = 0$; this condition directly corresponds to the strong shock limit of the Rankine-Hugoniot jump conditions, where the unshocked pressure and SIE are negligible compared to their shocked values. Moreover, the strong shock assumption eliminates the characteristic scales $p_0$ and $e_0$ from the problem formulation; the presence of any such scales usually reduces the inherent scaling symmetry rank, as discussed in Section~\ref{sec:gud_prob} (however, we will also show in Sec.~\ref{sec:shock_trajectory} that this rule must be applied with caution).

On the other hand, it may sometimes prove necessary to include a non-zero $p_0$ or $e_0$ in the formulation of a problem of interest. Consider, for example, a simple generalization of the ideal gas EOS provided by Eq.~(\ref{eqn:ideal_gas}),
\begin{equation}
  p \propto \rho (e-e_0) ,
  \label{eqn:ideal_gen}
\end{equation}
where $e_0$ is a reference SIE. With Eq.~(\ref{eqn:ideal_gen}), an unshocked $p_0=e_0=0$ state is only satisfied for $\rho_0=0$ -- a vacuum state. However, in this case an $e=e_0$ unshocked state allows for the satisfaction of $p_0=0$ for any unshocked density, and Eqs.~(\ref{eqn:jump_begin})-(\ref{eqn:jump_end}) still reduce to the strong shock limit. 

Since many EOS models are intended to only match experimental data at high pressures and energies, it is reasonable that situations such as the preceding might arise and demand flexibility with the unshocked conditions. Indeed, Schmidt et al.~\cite{emma} and Lilieholm et al.~\cite{jenni} have completed preliminary work to develop an EOS with a reference SIE feature similar to that included in Eq.~(\ref{eqn:ideal_gen}), which nevertheless exhibits Guderley-like solutions when used in conjunction with the inviscid Euler equations.

Thus, in this work, we assume $p_0=0$ but do not necessarily require $e_0=0$. We will show in Sec.~\ref{sec:shock_trajectory} that the $p_0=0$ restriction can be relaxed with very little change to the results, but we do not treat that case in full detail here.

\section{Existence theorem}
\label{sec:main}

The main result of this work is:
\begin{theorem}
All adiabatic bulk moduli for which the Guderley problem has a self-similar scaling solution satisfy $K_S=pf(\rho)$, where $f$ is an arbitrary function. In all such solutions, the shock is accelerating as $r \propto |t|^{1-\lambda}$ for $\lambda\in (0,1]$. The similarity exponent, $\lambda,$ if it exists, solves a nonlinear eigenvalue problem.
\end{theorem}
We now prove this result in three subsections. The first investigates the choice of similarity variables, the second analyzes in detail the jump conditions, shock trajectory, and unshocked conditions, and the third considers solvability. In a fourth subsection, we focus on a special case which seems not to have been previously considered, but which is nonetheless of great interest.

\subsection{Similarity variables}
\label{sec:derivation_of_variables}
\label{sec:sim_vars}

We work with the 1D cylindrically or spherically symmetric form of the inviscid Euler equations, given previously by Eqs.~(\ref{eqn:euler_spherical_begin})-(\ref{eqn:euler_spherical_end}),
\begin{eqnarray}
  \pd{\rho}{t} + u \pd{\rho}{r} + \rho\left( \pd{u}{r} + \frac{k u}{r} \right) = 0,  \label{eqn:cons_mass} \\
  \pd{u}{t}    + u \pd{u}{r}    + \frac{1}{\rho}\pd{p}{r} = 0, \label{eqn:cons_mom} \\
  \pd{p}{t}    + u \pd{p}{r}    + K_S\left( \pd{u}{r} + \frac{k u}{r} \right) = 0, \label{eqn:cons_eng}
\end{eqnarray}
and the self-similar scaling ansatz given previously by Eqs.~(\ref{eqn:sim_ansatz_begin})-(\ref{eqn:sim_ind}),
\begin{eqnarray}
  \rho(r,t) &=& |t|^{\beta_{\rho}} D(\xi), \label{eqn:sim_ansatz_begin_1} \\
  u(r,t) &=& |t|^{\beta_u} V(\xi), \label{eqn:sim_ansatz_mid_1} \\
  p(r,t) &=& |t|^{\beta_p} \Pi(\xi), \label{eqn:sim_ansatz_midd_1} \\
  \xi &=& r |t|^{-(1-\lambda)}. \label{eqn:sim_ansatz_end_1}
\end{eqnarray}
As discussed in Sec.~\ref{sec:gud_prob}, we seek values of $\beta_{\rho}$, $\beta_u$, $\beta_p$, and $\lambda$ such that Eqs.~(\ref{eqn:cons_mass})-(\ref{eqn:cons_eng}) collapse to ODEs in $\xi$ when the substitutions given by Eqs.~(\ref{eqn:sim_ansatz_begin_1})-(\ref{eqn:sim_ansatz_end_1}) are made. Before doing so, additional simplification can be achieved by examining the unshocked state,
\begin{eqnarray}
  \rho(r,t) &=& \rho_0 = \rm{const.}, \\
  u(r,t) &=& 0, \\
  p(r,t) &=& 0,
\end{eqnarray}
for $r<r_s(t)$ and all $t$. As a result, $\beta_{\rho} = 0$ appearing in Eq.~(\ref{eqn:sim_ansatz_begin_1}), since the constant unshocked density cannot scale as a function of time with $\xi$ held constant. There are no such constraints on $\beta_{u}$ and $\beta_{p}$ appearing in Eqs.~(\ref{eqn:sim_ansatz_mid_1}) and (\ref{eqn:sim_ansatz_midd_1}) since the $u$ and $p$ variables feature no characteristic scales in the unshocked region. 

Now, we apply Eqs.~(\ref{eqn:sim_ansatz_begin_1})-(\ref{eqn:sim_ansatz_end_1}) to Eq.~(\ref{eqn:cons_mass}) to yield 
\begin{equation}
  \left[ (1-\lambda) \xi + |t|^{\lambda+\beta_u} V \right]D' + |t|^{\lambda+\beta_u} \left(  V'+\frac{kV}{\xi} \right) = 0.
  \label{eqn:ode_1} 
\end{equation}
This equation depends on $\xi$ alone if either $\beta_u = -\lambda$ or $\lambda = 1$. The case $\lambda=1$ corresponds to a stationary shock, as will be shown in Sec.~\ref{sec:shock_trajectory}. Discarding this case and applying Eqs.~(\ref{eqn:sim_ansatz_begin_1})-(\ref{eqn:sim_ansatz_end_1}) to Eq.~(\ref{eqn:cons_mom}) yields
\begin{equation}
  \lambda V +(1-\lambda)\xi V' +VV' + \frac{\Pi'}{D}|t|^{\beta_p+2\lambda} = 0, 
  \label{eqn:ode_2}
\end{equation}
which depends only on $\xi$ if $\beta_p = -2\lambda$. As a result, Eqs.~(\ref{eqn:sim_ansatz_begin_1})-(\ref{eqn:sim_ansatz_end_1}) become
\begin{eqnarray}
  \rho(r,t) &=& D(\xi), \label{eqn:sim_final_begin} \\
  u(r,t) &=& |t|^{-\lambda} V(\xi),  \\
  p(r,t) &=& |t|^{-2\lambda} \Pi(\xi),  \\
  \xi &=& r |t|^{-(1-\lambda)}. \label{eqn:sim_final_end} 
\end{eqnarray}
Finally, applying Eqs.~(\ref{eqn:sim_ansatz_begin_1})-(\ref{eqn:sim_ansatz_end_1}) to Eq.~(\ref{eqn:cons_eng}) yields
\begin{equation}
  2\lambda \Pi + (1-\lambda)\xi \Pi' + V\Pi' +|t|^{2\lambda} K_S \left( V'+\frac{kV}{\xi} \right) = 0. 
  \label{eqn:ode_3}
\end{equation}
To cancel the dangling $t$ term from this relation, either $\lambda=0$, or $K_S$ must obey
\begin{equation}
  K_S \propto |t|^{-2\lambda}.
\end{equation}
Since $K_S$ can depend only on $p= |t|^{-2\lambda} \Pi(\xi)$ and $\rho = D(\xi)$, it then follows that
\begin{equation}
  K_S = p f(\rho),
  \label{eqn:magicK}
\end{equation}
where $f$ is an arbitrary function of the indicated argument. The derivation culminating in Eq.~(\ref{eqn:magicK}) is similar to those provided by Holm~\cite{holm1976symmetry}, Axford and Holm~\cite{axford_holm_1978}, Hutchens~\cite{hutchens}, and Boyd et al.~\cite{general_euler_symmetries} using symmetry analysis arguments. In any event, one of $\lambda=0$ or Eq.~(\ref{eqn:magicK}) are the necessary conditions to cancel the dangling $t$ appearing in Eq.~(\ref{eqn:ode_3}). (See Appendix~A for a formal derivation of this outcome.)

Equation~(\ref{eqn:magicK}) has a significant physical interpretation. For example, in the ideal gas case, $f(\rho)=\gamma$ as given by Eq.~(\ref{eqn:bmideal}) is a constant function. In any other case, we observe that the output of $f$ must be dimensionless [since $K_S$ has units of pressure, as may be ascertained from inspection of Eq.~(\ref{eqn:abm_def})]. This means there must be a characteristic density incorporated into $f$ that allows the units of $\rho$ to be canceled; this is the only characteristic scale permitted to exist in $f$. The reason that a characteristic density is allowed is by making the unshocked density fixed and non-zero, there is already one characteristic density present in the problem definition, so no symmetry is lost by incorporating another. Indeed, we have already seen this characteristic density destroy one degree of symmetry in Eq.~(\ref{eqn:sim_ansatz_begin_1}) when we were forced to exclude scaling on density in the construction of similarity variables.

Moving forward, we will assume either $\lambda=0$ or Eq.~(\ref{eqn:magicK}) holds, since the subsequent analysis is identical. It is notable that there is no restriction on $K_S$ in the case $\lambda=0$, so if a suitable solution of the associated ODEs can be found, it will apply to any flow that can be modeled by Eqs.~(\ref{eqn:cons_mass})-(\ref{eqn:cons_eng}) and an EOS closure model expressible in terms of $K_S$.

\subsection{Shock trajectory}
\label{sec:shock_trajectory}

It is customary in solving the Guderley problem to set the unshocked pressure and SIE equal to zero. We have already shown in Sec.~\ref{sec:unshocked} that this is not always possible, but without understanding the scaling law obeyed by Guderley solutions, we were unable to fully specify the correct boundary conditions or the shock trajectory. To do so, consider again the Rankine-Hugoniot shock jump conditions relevant to the Guderley problem,
\begin{align}
  \rho_1(u_s-u_1) &= \rho_0 u_s, \label{eqn:jump_again_begin} \\
  \rho_1 u_s u_1 &= p_1-p_0, \\
  \rho_0 u_s \left(e_1-e_0 + \frac{1}{2}u_{1}^2\right) &= p_1 u_1. \label{eqn:jump_again_end}
\end{align}
It is possible to assume a slightly more general self-similar scaling ansatz than given by Eqs.~(\ref{eqn:sim_ansatz_begin})-(\ref{eqn:sim_ansatz_end}), namely
\begin{eqnarray}
  \rho &=& |t|^{\beta_{\rho}} D(\xi), \\
  u &=& |t|^{\beta_u} V(\xi), \\
  p - p_0 &=& |t|^{\beta_p} \Pi(\xi),
\end{eqnarray}
where the independent similarity variable $\xi$ is still given by Eq.~(\ref{eqn:sim_ind}). This formulation yields a reduction of Eqs.~(\ref{eqn:cons_mass})-(\ref{eqn:cons_eng}) to ODEs provided we also have either $\lambda=0$ or
\begin{equation}
K_S = (p-p_0)f(\rho),
\end{equation}
and
\begin{eqnarray}
  \rho &=& D(\xi), \label{eqn:sim_ansatz_gen_begin} \\
  u &=& |t|^{-\lambda} V(\xi),  \\
  p-p_0 &=& |t|^{-2\lambda} \Pi(\xi),  \\
  \xi &=& r |t|^{-(1-\lambda)}. \label{eqn:sim_ansatz_gen_end}
\end{eqnarray}
The derivation of this slight generalization proceeds identically to the presentation provided in Sec.~\ref{sec:sim_vars},\footnote{This form appears explicitly in the work of Axford and Holm~\cite{axford_holm_1978} and is implicitly present in the works of Ovsiannikov~\cite{ovsiannikov_book} and Boyd et al~\cite{general_euler_symmetries}. Although Axford and Holm show that the similarity variable $p-p_0 = |t|^{-2\lambda}\Pi(\xi)$ can lead to a reduction to ODEs, it appears that the example in their work uses $p_0=0$. Thus, to the knowledge of the authors, there is no example where this more general case is actually applied. Moreover, the unshocked conditions associated with $p_0>0$ seem not to have been explicitly analyzed in other works.}.

As noted in Sec~\ref{sec:unshocked}, the assumption $p_0=e_0=0$ is not always thermodynamically consistent for a particular choice of $K_S$. Instead, we propose setting $\rho_0$ and $p_0$ and let $e_0$ be whatever value is dictated by the EOS. We now show that these choices lead to jump conditions that depend on $\xi$ alone, thus preserving the self-similar nature of the Guderley problem. Inserting Eqs.~(\ref{eqn:sim_ansatz_gen_begin})-(\ref{eqn:sim_ansatz_gen_end}) into Eqs.~(\ref{eqn:jump_again_begin})-(\ref{eqn:jump_again_end}) gives
\begin{align}
  D_1(u_s-V_1 |t|^{-\lambda}) &= u_s, \label{eqn:jump_sim_begin} \\
  D_1 u_s V_1 |t|^{-\lambda} &= |t|^{-2\lambda} \Pi_1, \label{eqn:jump_sim_mid} \\
  \rho_0 u_s \left(e_1-e_0 + \frac{1}{2}V_1{^2} |t|^{-2\lambda} \right) &= V_1 \Pi_1 |t|^{-3\lambda}. \label{eqn:jump_sim_end}
\end{align}
In order for these conditions to be expressible solely in terms of similarity variables, from inspection of Eq.~(\ref{eqn:jump_sim_begin}) it is necessary that the shock velocity assume the form
\begin{equation}
  u_s \propto |t|^{-\lambda} ,
  \label{eqn:shock_vel}
\end{equation}
whence
\begin{equation}
  r_s \propto |t|^{1-\lambda}.
  \label{eqn:shock_pos}
\end{equation}
With Eq.~(\ref{eqn:sim_final_end}), Eq.~(\ref{eqn:shock_pos}) indicates that the converging shock trajectory exists along a constant value $\xi_s$ in $\xi$-space, so that, more preceisely,
\begin{eqnarray}
  r_s &=& \xi_s |t|^{1-\lambda}, \label{eqn:shock_pos_2} \\
  u_s &=& -(1-\lambda) \xi_s |t|^{-\lambda}. \label{eqn:shock_vel_2}
\end{eqnarray}
Equations~(\ref{eqn:shock_pos_2}) and (\ref{eqn:shock_vel_2}) then show that $\xi_s >0$ and $\lambda\in (0,1]$ are necessary conditions for the shock wave to accelerate toward $r=0$ as $t \to 0$.

Moreover, with Eqs.~(\ref{eqn:shock_pos_2}) and (\ref{eqn:shock_vel_2}), Eqs.~(\ref{eqn:jump_sim_begin})-(\ref{eqn:jump_sim_end}) become
\begin{align}
  D_1\left[ (1-\lambda)\xi_s+V_1 \right] &= (1-\lambda) \xi_s, \label{eqn:jump_sim_again_begin} \\
  -(1-\lambda) D_1 \xi_s V_1 &= \Pi_1, \\
  -(1-\lambda) D_0 \xi_s \left(e_1-e_0 + \frac{1}{2}V_1{^2} |t|^{-2\lambda} \right) &= V_1 \Pi_1 |t|^{-2\lambda} + V_1 p_0.
  \label{eqn:jump_sim_again_end}
\end{align}
At this point, it would appear that Eq.~(\ref{eqn:jump_sim_again_end}) is irreducible to similarity variables alone if $e_0$ and $p_0$ are not zero. This is not quite true; with Eq.~(\ref{eqn:Ks_invert}) Boyd et al.~\cite{general_euler_symmetries} show that $e$ is necessarily of the form
\begin{equation}
e = (p-p_0)g(\rho) + \frac{p_0}{\rho} + h\left( \frac{p-p_0}{\rho^2 g'(\rho)} \right), 
\end{equation}
where $g$ and $h$ are arbitrary functions of their arguments (the latter corresponding to the kernel of a linear PDE solution), and $p_0/\rho$ may be interpreted as a reference SIE. It appears that $h$ corresponds to adding a constant to each adiabat -- in the following, we set $h=0$ for simplicity, and also because the corresponding term in the ideal gas case has $h=0$, as shown by Boyd et al~\cite{general_euler_symmetries}. Thus, we assume
\begin{equation}
e - \frac{p_0}{\rho} = (p-p_0)g(\rho),
\end{equation}
for the arbitrary function $g$. Substituting this additional information into Eq.~(\ref{eqn:jump_sim_again_end}) then yields
\begin{equation} 
-(1-\lambda) D_0 \xi_s \left(|t|^{-2\lambda} \Pi + 	\frac{p_0}{D_1}  - 0 - \frac{p_0}{D_0} + \frac{1}{2}V_1{^2} |t|^{-2\lambda} \right) = V_1 \Pi_1 |t|^{-2\lambda} + V_1 p_0,
\label{eqn:jump_sim_energy}
\end{equation}
and from Eq.~(\ref{eqn:jump_sim_again_begin}) we have
\begin{equation}
V_1 = - D_0 \xi_s \left( \frac{1}{D_0} - \frac{1}{D_1} \right), 
\end{equation}
so that Eq.~(\ref{eqn:jump_sim_energy}) becomes
\begin{equation}
-(1-\lambda) D_0\xi_s g(D_1) = V_1, 
\end{equation}
which is a relation depending only on $\xi$. Converting this relation to physical variables yields
\begin{equation}
\frac{u_1}{u_s} = \rho_0 g(\rho_1).
\end{equation}
This relation shows that if $K_S$ is of the form $(p-p_0) f(\rho)$ (with the unshocked pressure given by $p_0$), then we can expect self-similar scaling to occur, even in the presence of a characteristic unshocked pressure and energy. As a result, the conventional wisdom concerning characteristic scales and self-similar scaling must be applied with caution: a Guderley-like problem may still exist for EOS closures that do not admit the standard initial conditions $p_0=e_0=0$.\footnote{If one instead chooses the standard ansatz $p = |t|^{-2\lambda}\Pi(\xi)$, it is still possible to have a non-zero unshocked energy, but it is very difficult to have a nonzero unshocked pressure, as calculations very similar to the foregoing show.}

It is worth noting that there is another approach that allows for non-zero unshocked pressures and energies, namely that used in the literature on finite-strength shocks (see, for example, Sedov~\cite{sedov_book} or Hutchens~\cite{hutchens}). The details of such an analysis are outside the scope of this work, but we note that it does not yield exact self-similarity. Rather, it yields an approximate self-similarity or ``quasi-similarity,'' as discussed at length by the aforementioned authors and, for example, Oshima~\cite{oshima1960blast}, Lee~\cite{lee1967nonuniform}, Rae~\cite{rae1970analytical}, Axford and Holm~\cite{axford1981converging}, and Hafner~\cite{hafner}.

It is also worth noting that a Guderley-like problem with a spatially-variable unshocked density is considered by Lazarus \cite{lazarus}, Meyer-ter-Vehn and Schalk~\cite{meyer_ter_vehn_schalk}, Toque~\cite{toque}, and Madhumita~\cite{madhumita}.

\subsection{Analysis of ODEs}
\label{sec:analysis_odes}

Equations~(\ref{eqn:ode_1}), (\ref{eqn:ode_2}), and (\ref{eqn:ode_3}) may be rewritten as
\begin{eqnarray}
  X D' + DV' &=& -\frac{k}{\xi}DV, \label{eqn:p_linear_odes_begin} \\
  X V' + \frac{1}{D}\Pi' &=& -\lambda V, \\
  X \Pi' + K_S |t|^{2\lambda}\left( V' + \frac{k}{\xi}V \right) &=& -2\lambda \Pi,
  \label{eqn:p_linear_odes_end}
\end{eqnarray}
where we have substituted 
\begin{equation}
  X=(1-\lambda)\xi+V,
  \label{eqn:X_def}
\end{equation}
for notational brevity. As noted in Sec.~\ref{sec:sim_vars}, this system collapses to ODEs only if $\lambda=0$ or Eq.~(\ref{eqn:magicK}) is satisfied. Assuming one of these conditions, isolating the derivatives in Eqs.~(\ref{eqn:p_linear_odes_begin})-(\ref{eqn:p_linear_odes_end}) gives the system
\begin{equation}
  \left(\begin{array}{c}
    D' \\
    V' \\
    \Pi' \\
  \end{array}\right) = \frac{-V}{D\left( X^2 - C^2 \right)}
  \left(
  \begin{array}{c}
    D\left( k \frac{X}{\xi}-\lambda \right)+\frac{2\lambda\Pi}{XV}\\
    \lambda\left( X-\frac{2\Pi}{DV} \right)-\frac{k C^2}{\xi} \label{eqn:doesnt_vanish}\\
    DC^2\left( k \frac{X}{\xi}-\lambda \right)+2\lambda \Pi\frac{X}{V}
  \end{array}
  \right)
\end{equation}
where $C(\xi) \equiv |t|^\lambda c$ may be interpreted as a scaled sound speed that replaces $K_S$ via Eq.~(\ref{eqn:sound_speed_def}) -- this substitution helps provide a more intuitive physical interpretation to the denominator appearing in Eq.~(\ref{eqn:doesnt_vanish}). This interpretation will be important in the subsequent analysis.

Consistent with the formal definition of the Guderley problem provided in Section~\ref{sec:gud_prob}, we seek a solution of Eq.~(\ref{eqn:doesnt_vanish}) that is \emph{everywhere bounded} in the variables $D$, $V$, and $\Pi$ (a direct consequence of the boundedness condition on the physical variables $\rho$, $u$, and $P$). Since we assume $K_S$ is bounded, by its definition $C$ is also bounded. We are also only interested in differentiable solutions of Eq.~(\ref{eqn:doesnt_vanish}), and will discard all choices of $\lambda$ and $C$ that cause $D'$, $V'$, and $\Pi'$ to become infinite. We therefore analyze whether or not the denominator appearing in Eq.~(\ref{eqn:doesnt_vanish}) vanishes. Since this denominator is a continuous function of its arguments, it will vanish if we can identify one coordinate where it is negative and another where it is positive.

Far from the origin (i.e., $r \to \infty$) and for all finite times, $\xi \to \infty$ by Eq.~(\ref{eqn:sim_ind}). As a result, the denominator of Eq.~(\ref{eqn:doesnt_vanish}), or
\begin{equation}
  D \left( X^2 - C^2 \right) = D \left[ \left( 1-\lambda \right) \xi + V \right]^2 -DC^2,
  \label{eqn:denom_odes}
\end{equation}
is observed by inspection to be strictly positive at the aforementioned state, as all quantities besides $\xi$ appearing in it are bounded.  

Immediately adjacent to the shock front, the denominator is negative. To see this, observe that if $\xi_s$ is the position of the shock wave in $\xi$-space, then the physical shock trajectory $r_s$ and shock velocity $u_s$ are given by Eqs.~(\ref{eqn:shock_pos_2}) and (\ref{eqn:shock_vel_2}), respectively. Accordingly,
\begin{equation}
  \xi_s = \frac{-u_s |t|^{\lambda}}{1-\lambda},
\end{equation}
so that with Eq.~(\ref{eqn:X_def}), it then follows that 
\begin{equation}
  X = |t|^{\lambda}(u-u_s),
  \label{eqn:X_shock}
\end{equation}
immediately adjacent to the shock front. The quantity $u-u_s$ appearing in Eq.~(\ref{eqn:X_shock}) is the shock velocity in a reference frame where the post-shock particles are motionless. With Eq.~(\ref{eqn:X_shock}) and the definition of $C$, the denominator appearing in Eq.~(\ref{eqn:doesnt_vanish}) may be written in terms of physical variables as 
\begin{equation}
  \rho |t|^{2\lambda} \left[ (u_s-u)^2-c^2 \right].
  \label{eqn:denom_shock}
\end{equation}
Thus, to determine the sign of this quantity when evaluated immediately adjacent to the shock front, we need only determine whether the shocked sound speed is bigger or smaller than the aforementioned motionless-particle reference frame shock velocity. The conclusion follows exactly from the shock stability assumption provided in Section~\ref{sec:gud_prob}: we assume via thermodynamic and perturbative stability arguments that the shock travels subsonically relative to a particle immediately behind it. As a result, Eq.~(\ref{eqn:denom_shock}) is strictly negative when evaluated immediately adjacent to the shock front.

Combined with Eq.~(\ref{eqn:denom_odes}) evaluated as $\xi \to \infty$, this result ensures that the denominator appearing in Eq.~(\ref{eqn:doesnt_vanish}) does indeed change sign between the shock front and $r \to \infty$, and will somewhere vanish. Thus, Eq.~(\ref{eqn:doesnt_vanish}) does not have a bounded, differentiable solution, except if the numerators appearing within can all be made to vanish simultaneously with the denominator. Since by the definition of the Guderley problem none of the variables $D$, $V$, $\Pi$, and $C$ can be made zero, only the parameter $\lambda$ may be selected to (in principle) achieve the required condition. This result shows that the counterpart phenomenon appearing in the ideal gas Guderley solution is observed in general: Eq.~(\ref{eqn:doesnt_vanish}) is a nonlinear eigenvalue problem. In the ideal gas case, the correct value of $\lambda$ is determined numerically, as has been done among many others by Guderley~\cite{guderley}, Stanyukovich~\cite{stanyukovich}, Butler~\cite{butler}, Lazarus~\cite{lazarus}, Chisnell~\cite{chisnell}, and Ramsey et al~\cite{guderley_revisited}.

The preceding proof is notable in that it gives an intuitive meaning to the blowup observed in solving Eq.~(\ref{eqn:doesnt_vanish}): the similarity variable derivatives are proportional to the inverse difference between the sound speed and the shock speed in a certain reference frame, at least immediately adjacent to the shock front: 
\begin{equation}
D',V',\Pi' \propto \left[ (u_s-u)^2-c^2 \right]^{-1}.
\end{equation}
Adjacent to the shock front, stability requires that this difference have one sign, whereas far from the shock, boundedness requires it to have the opposite sign. Thus, one is forced to either abandon boundedness (as done by Velikovich et al.~\cite{mhd} and Coggeshall~\cite{coggeshall1991analytic} in some shock-free solutions) or solve an eigenvalue problem in the style of Guderley and his many successors. As far as the authors are aware, this interpretation is novel; the complete theory for when the eigenvalue problem has a solution (or what the solution might be in general) is left as a matter for future work.

\subsection{The universal symmetry}
\label{sec:lambda_zero}

In Section~\ref{sec:analysis_odes} the reduction to ODEs was accomplished by assuming either $\lambda=0$ or $K_S=pf(\rho)$. While the latter case notably includes the ideal gas Guderley and related problems as members, the $\lambda=0$ case is much broader, encompassing all materials modeled by Eqs.~(\ref{eqn:cons_mass})-(\ref{eqn:cons_eng}) with an EOS closure model expressible in terms of an adiabatic bulk modulus. If $\lambda=0$ solutions can be found, they yield code verification test problems for a broad class of flows and materials. This case has not been studied rigorously in the context of the Guderley solution, although it has been treated by Bethe~\cite{bethe}, Menikoff and Plohr~\cite{menikoff_and_plohr}, Kamm~\cite{kamm_riemann}, and Ramsey et al.~\cite{our_shock_waves_paper} for various Riemann and Noh problems.

With Eq.~(\ref{eqn:sim_final_end}), setting $\lambda=0$ gives the similarity variable
\begin{equation}
  \xi = \frac{r}{|t|},
\end{equation}
which, following the arguments presented in Sec.~\ref{sec:shock_trajectory}, implies constant velocity shock waves following the trajectories
\begin{equation}
  r_s =\xi_s t.
\end{equation}
With this parameterization, the ODEs given by Eq.~(\ref{eqn:doesnt_vanish}) become
\begin{equation}
  \left(\begin{array}{c}
    D' \\
    V' \\
    \Pi' \\
  \end{array}\right) = \frac{-V}{D\left[ X^2 - C^2 \right]}
  \left(
  \begin{array}{c}
    \frac{kXD}{\xi} \\
    -\frac{k C^2}{\xi} \label{eqn:doesnt_vanish_0} \\
    \frac{kXDC^2}{\xi} 
  \end{array}
  \right) ,
\end{equation}
where Eq.~(\ref{eqn:X_def}) likewise becomes
\begin{equation}
  X=\xi+V .
\end{equation}
The singular analysis of Eq.~(\ref{eqn:doesnt_vanish_0}) proceeds identically to that of Eq.~(\ref{eqn:doesnt_vanish}) presented in Sec.~\ref{sec:analysis_odes}, with identical conclusions: the denominator of Eq.~(\ref{eqn:doesnt_vanish_0}) is strictly positive as $r \to \infty$, and strictly negative at $r = r_s(t)$. As before, the denominator of Eq.~(\ref{eqn:doesnt_vanish_0}) must therefore vanish somewhere, resulting in an unbounded solution contrary to the definition of the Guderley problem as outlined in Sec.~\ref{sec:gud_prob}.

However, for $\lambda=0$ the zero-denominator pathology cannot be averted simply by requiring the numerators of Eq.~(\ref{eqn:doesnt_vanish_0}) to simultaneously vanish with its denominator: again, none of $D$, $V$, $\Pi$, and $C$ may be made zero, as $\rho$, $u$, and $c$ must be non-zero and finite for all $t<0$, by the definition of the Guderley problem. Thus, by requiring $\lambda=0$, we have ostensibly removed a degree of freedom that would otherwise allow for at least the possibility of a nonlinear eigenvalue problem, and thus the only route to the construction of a bounded solution.\footnote{In the scoping study of Lilieholm et al.~\cite{jenni}, it appears there is a global, unbounded solution of Eq.~(\ref{eqn:doesnt_vanish_0}) in some cases, which is a promising avenue for future work.} In return, we have gained one more degree of freedom in the choice of $K_S$, but the above analysis shows that this freedom is not enough to generate any Guderley-like solutions. One potential path to increasing the utility of the $\lambda=0$ case is to investigate it in conjunction with flows featuring multiple shock waves or other discontinuities, as discussed by Lazarus~\cite{lazarus}. 

Otherwise, the restriction on EOS closure models satisfying $K_S=pf(\rho)$ is definitively required in all cases.

\section{Conclusion}
\label{sec:conclusion}

In the foregoing, we have analyzed in detail the Guderley problem for an arbitrary EOS closure model by showing how to set up the initial conditions, calculating the motion of the shock wave, determining all possible reductions to similarity variables, and showing that the resulting ODE system and boundary conditions constitute a nontrivial eigenvalue problem. In particular, we have shown that the ``universal'' choice of similarity variables always leads to a reduction to ODEs, never conflicts with the boundary conditions, and yet yields ODEs that never have a bounded, differentiable solution. This motivates further study of when the ODEs resulting from the introduction of similarity variables have a solution with reasonable properties. For instance, from the foregoing, it is clear that one must choose between unbounded solutions or an eigenvalue problem. Another novel feature of this work is that, by conducting our analysis of the governing ODEs on a more abstract level than is usually done, we have arrived at an intuitive explanation for the necessity of solving an eigenvalue problem to obtain a self-similar solution; namely, there is a competition between thermodynamic stability of the shock wave and boundedness of the shocked driving conditions.

\section*{Acknowledgements}

This work was performed under the auspices of the United States Department of Energy by Los Alamos National Security, LLC, at Los Alamos National Laboratory under contract DE-AC52-06NA25396. Z. Boyd was also supported by an NDSEG Fellowship. The authors thank J. Schmidt, E. Schmidt, J. Lilieholm, W. Black, and J. Ferguson for helpful conversations and valuable insights on these topics.

\nocite{*}
\bibliographystyle{my_qjmam}
\bibliography{boyd_guderley}

\begin{appendices}

\section{Formal derivation of the scale-invariant $K_S$}

A more formal derivation of Eq.~(\ref{eqn:magicK}) proceeds along these lines: suppose there exists a self-similar scaling solution to the Guderley problem for a particular $K_S$. Such a solution yields functions $D,\Pi:(0,\infty)\to [0,\infty)$ that are well-defined. Moreover, as outlined in Sec.~\ref{sec:sim_vars}, the function
\begin{equation}
F(\xi,t) = |t|^{2\lambda} K_S\left[ |t|^{-2\lambda} \Pi(\xi),D(\xi) \right],
\label{eqn:appx_1}
\end{equation}
depends only on $\xi$, or,
\begin{equation}
F(\xi,t) = F(\xi,t_0),
\end{equation}
for any choice of $\xi, t$, and $t_0$. Let $t_0$ be fixed; expanding Eq.~(\ref{eqn:appx_1}) then gives
\begin{equation}
|t|^{2\lambda} K_S\left[ |t|^{-2\lambda} \Pi(\xi), D(\xi) \right] = |t_0|^{2\lambda} K_S\left[ |t_0|^{-2\lambda} \Pi(\xi), D(\xi) \right].
\end{equation}
Rearranging, we have
\begin{equation} 
K_S\left[ |t|^{-2\lambda} \Pi(\xi),D(\xi) \right] = y(\xi) |t|^{-2\lambda}, 
\end{equation}
where $y$ is an arbitrary function of the indicated argument.

Now, let $a>0$ be a constant, and let $x_2$ be an arbitrary element of the range of $D$ [i.e., $x_2 = D(\xi)$]. Then, for any $x_1>0$, we can choose $t\in (-\infty,0)$ so that $x_1 = |t|^{-2\lambda} \Pi(\xi)$, since $\Pi(\xi)$ is assumed to be positive. Then, we have
\begin{eqnarray}
    K_S(\alpha x_1,x_2) &=& K_S\left[\alpha |t|^{-2\lambda} \Pi(\xi),D(\xi)\right] \nonumber \\
    &=& K_S\left[ \left|a^{\frac{-1}{2\lambda}}t\right|^{-2\lambda}\Pi(\xi),D(\xi) \right] \nonumber \\
    &=& c(\xi) \left|a^{\frac{-1}{2\lambda}} t \right|^{-2\lambda} \nonumber \\
    &=& a c(\xi) |t|^{-2\lambda} \nonumber \\
    &=& a K_S\left[ |t|^{-2\lambda} \Pi(\xi) , D(\xi) \right] \nonumber \\
    &=& a K_S(x_1,x_2),
\end{eqnarray}
and thus, $K_S$ is homogeneous in its first argument (at least in the domain needed for the Guderley problem). 

\end{appendices}

\end{document}